
entire tex paper, standard tex, no figures

\magnification=1200
\parindent=0pt
\vskip0.5in
\bf
\centerline{SUM RULE OF THE CORRELATION FUNCTION}
\rm
\vskip0.5in
\centerline{Stanis\l aw Mr\' owczy\' nski\footnote{*}{E-mail
address: MROW@PLEARN.bitnet}}
\vskip0.1in
\centerline{High-Energy Department}
\centerline{Soltan Institute for Nuclear Studies}
\centerline{ul. Ho\. za 69, PL - 00-681 WARSAW, Poland}
\vskip1in

\baselineskip=21pt

ABSTRACT: We derive a sum rule satisfied by the correlation function
of two particles with small relative momenta, which results from the
completeness condition of the quantum states.

\vskip1in

\centerline{\it to appear in Physics Letters B}

\vskip0.3in

\centerline{\it November 1994}

\parindent=20pt

\vfill
\eject

The correlation functions of two identical or nonidentical particles with
`small' relative momenta have been extensively studied in nuclear collisions
for bombarding energies from tens of MeV [1] to hundreds of GeV [2].
These functions provide unique information about space-time characteristics
of particle sources in the collisions. We show in this paper that the
correlation function integrated over particle relative momentum satisfies
a simple relation due to the completeness of the particle quantum
states. The preliminary account of this work has been presented in [3].

The correlation function ${\cal R}$ is defined as
$$
{ d n \over d{\bf p}_1 d {\bf p}_2 }= {\cal R}({\bf p}_1,{\bf p}_2)
{ d n \over d{\bf p}_1 } { d n \over d {\bf p}_2 } \;,
$$
where $ d n / d{\bf p}_1 d {\bf p}_2 $ and $d n / d{\bf p}_1 $
is the two- and one-particle momentum distribution normalized to unity.
It has been repeatedly argued [4] that the correlation function ${\cal R}$
can be expressed in the source rest frame in the following way
$$
{\cal R}({\bf p}_1,{\bf p}_2) = \int d^3r_1 dt_1  \int d^3r_2 dt_2 \;
{\cal D}({\bf r}_1,t_1) {\cal D}({\bf r}_2,t_2) \;
\vert \psi ({\bf r}_1 ',{\bf r}_2 ') \vert ^2
 \;,  \eqno(2)
$$
where the source function ${\cal D}({\bf r},t)$, which is  normalized as
$\int d^3r {\cal D}({\bf r},t) = 1$, gives the probability to emit a nucleon
from a space-time point $({\bf r},t)$\footnote{*}{It should be understood
here that the coordinates $({\bf r},t)$ determine the position of a particle
wave-package center.}; $\psi $ is the final state wave function of the pair;
${\bf r}_i'\equiv {\bf r}_i -{\bf v}_i t_i$, $i = 1,2$ with ${\bf v}_i$
being the particle velocity relative to the source.

Eq. (2) determines the two particle correlation function as an overlap
of the source function and the final state wave function squared of
the two particles. The pair is assumed to be isolated from the rest
of the system not only in the final state, but starting from the moment
of {\it emission} or {\it freeze-out} when, due to the system decay
or expansion, the (strong) interaction is switched off\footnote{**}{
The pair cannot be treated as isolated even for the pair, which does
not interact with the rest of the system, in the case {\it many} identical
particles. Then, the wave function of the system does not factorize into
the pair wave function and the wave function of the rest, since the complete
wave function must be (anti-)symmetrized with respect to all particles.
However, the effect is significant only when the density  of the identical
particles is {\it large}. It  does not happen at the currently available
energies of nuclear collisions [5].}. (If the long range Coulomb force
is important after the `strong' freeze-out, the pair motion in the external
electromagnetic field should be considered [6].) The overlap from eq. (2)
is computed at the freeze-out, which for the pair equals $max(t_1,t_2)$,
and then is averaged over $t_1$ and $t_2$. Thus, the correlation function
carries the information on the system only at the moment of freeze-out and
not at earlier times when the pair of the particles still interacts with
the rest system.

Since we are interested in the correlations of particles with `small'
relative momenta, one can factorize the center-of-mass and relative
motion of the two particles in the essentially nonrelativistic manner.
Then, after eliminating the center-of-mass motion, eqs. (2) can be
rewritten as
$$
{\cal R}({\bf q}) = \int d^3r dt  \; {\cal D}_r ({\bf r},t) \;
\vert \phi_{\bf q}({\bf r}') \vert ^2
\;,  \eqno(3)
$$
with $ {\cal D}_r ({\bf r},t)$ being the distribution of the relative
space-time position of the two particles,
$$
{\cal D}_r ({\bf r},t) =
\int d^3R \; dT  \;
{\cal D}({\bf R}+{\bf r}/2,T+t/2) \;
{\cal D}({\bf R}-{\bf r}/2,T-t/2) \;.
$$
$\phi_{\bf q}({\bf r}') $ is the nonrelativistic wave function of
the relative motion with ${\bf q}$ denoting the particle momentum in
the center-of-mass frame of the pair. While the particle relative
motion is nonrelativistic, the center-of-mass motion with respect
to the source is, in general, relativistic. Therefore, the particle
relative distance measured in their center-of-mass frame ${\bf r}'$
is obtained by means of the Lorentz transformation i.e.
$$
{\bf r}' =  {\bf r} + (\gamma - 1) ({\bf r}{\bf n}) {\bf n}
- \gamma {\bf v} t \;,
\eqno(4)
$$
where {\bf v} is the pair velocity with respect to the source,
${\bf n} \equiv {\bf v} / \vert {\bf v} \vert$ and $\gamma$ is
the Lorentz factor of the center-of-mass motion relative to the source.
The correlation function (3) also depends on the total momentum of
the pair. This dependence, which is irrelevant for our considerations,
is not shown up.

Let us consider the correlation function integrated over
the relative momentum. Since ${\cal R}({\bf q}) \rightarrow 1$ when
${\bf q} \rightarrow \infty$, we rather discuss the integral
of ${\cal R}({\bf q}) - 1$. Using eq. (3) one immediately finds
$$
\int {d^3 q \over (2\pi )^3} \; \Big( {\cal R}({\bf q}) - 1 \Big)
= \int d^3r dt \;  {\cal D}_r ({\bf r},t) \;
\int {d^3 q \over (2\pi )^3} \;
\Big( \vert \phi_{\bf q}({\bf r}') \vert ^2 - 1 \Big)
\;.  \eqno(5)
$$

The wave functions satisfy the completeness condition
$$
\int {d^3 q \over (2\pi )^3} \;
\phi_{\bf q}({\bf r}) \phi^*_{\bf q}({\bf r}') +
\sum_{\alpha} \phi_{\alpha}({\bf r}) \phi^*_{\alpha}({\bf r}')
= \delta^{(3)}({\bf r} - {\bf r}') \pm \delta^{(3)}({\bf r} + {\bf r}')
\;,  \eqno(6)
$$
where $\phi_{\alpha}$ represents a bound state of the two particles.
When the particles are not identical the second term in the r.h.s of
eq. (6) should be neglected. This term guarantees the right symmetry
of both sides of the equation for the case of identical particles.
The upper sign is for bosons while the lower one for fermions.
The wave function of identical bosons (fermions) $\phi_{\bf q}({\bf r})$
is (anti-)symmetric when ${\bf r} \rightarrow -{\bf r}$, and the r.h.s of
eq. (6) is indeed (anti-)symmetric when ${\bf r} \rightarrow -{\bf r}$
or ${\bf r}' \rightarrow -{\bf r}'$. If the particles of interests carry
spin, the summation over the spin degrees of freedom in the l.h.s of
eq. (6) is implied.

When the integral representation of $\delta^{(3)}({\bf r} - {\bf r}') $
is used, eq. (6) can be rewritten as
$$
\int {d^3 q \over (2\pi )^3} \; \Big(
\phi_{\bf q}({\bf r}) \phi^*_{\bf q}({\bf r}') -
e^{i{\bf q}({\bf r}-{\bf r}')} \Big) +
\sum_{\alpha} \phi_{\alpha}({\bf r}) \phi^*_{\alpha}({\bf r}')
= \pm \delta^{(3)}({\bf r} + {\bf r}')
\;.
$$
Now we take the limit ${\bf r} \rightarrow {\bf r}'$ and get the relation
$$
\int {d^3 q \over (2\pi )^3} \;
\Big( \vert \phi_{\bf q}({\bf r}') \vert ^2 - 1 \Big)
=  \pm \; \delta^{(3)}(2 {\bf r}')
- \sum_{\alpha} \vert \phi_{\alpha}({\bf r}') \vert ^2
\;.  \eqno(7)
$$

When eq. (7) is substituted into eq. (5), we get the desired sum rule
$$
\int d^3 q \; \Big( {\cal R}({\bf q}) - 1 \Big)
= \pm {\pi^3 \over \gamma} \int dt \; {\cal D}_r ({\bf v}t/\gamma ,t)
- \sum_{\alpha} {\cal A}_{\alpha}
\;,  \eqno(8)
$$
where ${\cal A}_{\alpha}$ is the formation rate of a bound state
${\alpha}$ [7]
$$
{\cal A}_{\alpha} =  (2\pi)^3 \int d^3r dt \;
{\cal D}_r({\bf r},t) \vert \phi_{\alpha}({\bf r}') \vert ^2
 \;,
$$
which connects the cross section to produce the bound state ${\alpha}$
carrying the momentum ${\bf P}$ with that one of the two particles
with the momenta ${\bf P}/2$ as
$$
{d \sigma ^{\alpha} \over d{\bf P}} = \gamma  {\cal A}_{\alpha} \;
{ d \;\widetilde \sigma  \over d({\bf P}/2) d({\bf P}/2) }
\;.
$$
The tilde means that the short range correlations are removed from
the two-particle cross section, which is usually taken as a product
of the single-particle cross sections.

If the particles are emitted simultaneously (more precisely,
if $\langle {\bf r}^2 \rangle \gg \langle {\bf v}^2 t^2 \rangle $)
the source function is expressed as
${\cal D}_r ({\bf r},t)$ = ${\cal D}_r ({\bf r})\;\delta(t)$, and
the sum rule simplifies to
$$
\int d^3 q \; \Big( {\cal R}({\bf q}) - 1 \Big)
= \pm {\pi^3 \over \gamma} \;  {\cal D}_r (0)
- \sum_{\alpha} {\cal A}_{\alpha}
\;.
$$

The completeness condition is, obviously, valid for any inter-particle
interaction. It is also valid when the pair of particles interact
with the time-independent external field, e.g. the Coulomb field,
generated by the particle source. Thus, the sum rule (8) holds under
very general conditions as long as the basic formula (2) is justified,
in particular as long as the source function ${\cal D}_r ({\bf r},t)$
is ${\bf q}-$independent and spin independent. The validity of these
assumptions can be only tested within a microscopic model of nucleus--nucleus
collision. Below we consider three examples of the sum rule (8).

\vskip0.2in
{\it 1) The correlation function of identical pions.}
In this case the sum rule reads
$$
\int d^3 q \; \Big( {\cal R}_{\pi \pi} ({\bf q}) - 1 \Big)
= \lambda \; {\pi^3 \over \gamma} \int dt \;
{\cal D}_r ({\bf v}t/\gamma ,t) \;,
\eqno(9)
$$
where we have introduced {\it ad hoc} the chaoticity parameter $\lambda$.
As well known, the interferometric formula (2) gives $\lambda = 1$
in conflict with the experimental data which provide $\lambda <1$.
The sum rule (9) was earlier found by Podgoretzky [9] who used the free
wave function of pions and then explicitly integrated the correlation
function.

The relation (9) is approximately satisfied by the experimental
correlation function. The point is that the data are well described by
the free wave functions of the two pions (with the Coulomb correction
included) [2], which form the complete set of the quantum states.

\vskip0.2in
{\it 2) The p-p and n-n correlation function.} The sum rule (8) for
the identical nucleons is
$$
\int d^3 q \; \Big( {\cal R}_{NN} ({\bf q}) - 1 \Big)
= - {\pi^3 \over \gamma} \int dt \;
{\cal D}_r ({\bf v}t/\gamma ,t) \;.
\eqno(10)
$$
This relation, which, in particular, predicts exactly the same (negative)
value of the integral of the n-n and p-p correlation function, is {\it not}
satisfied by the experimental data [8]. The reason is probably the following.

The integration over ${\bf q}$ runs in the sum rule (10) to infinity.
Thus even a small deviation of ${\cal R}_{NN} ({\bf q}) $
from unity at `large' ${\bf q}$ can provide a sizeable contribution
to the integral (10). On the other hand, it is a serious problem
to normalize the experimental correlation function, and one usually
assumes that ${\cal R}_{NN} ({\bf q}) = 1$ at `large' ${\bf q}$.
Consequently, the sum rule (10) can be then easily violated.

\vskip0.2in
{\it 3) The n-p correlation function.} The sum rule (8) now reads
$$
\int d^3 q \; \Big( {\cal R}_{np}({\bf q}) - 1 \Big)
= - {\cal A}_{d}
\;,
$$
where the correlation function is averaged over spin. As expected,
the number of correlated n-p pairs is directly related to the
number of the produced deuterons. As in the previous case the
sum rule is {\it not} satisfied by the data [8], and the reason
is presumably the same.

\vskip0.2in

We conclude our considerations as follows.
Due to the completeness of the quantum states, the correlation functions
satisfy the simple relation which, in particular, connects the number
of correlated neutron-proton pairs with the number of deuterons produced
in nuclear collisions. It appears difficult to apply the sum rule to
the experimental data, however the relation is useful, at least, to test
theoretical calculations.

\vskip0.2in

{\it I am grateful to P. Danielewicz, A. Deloff, V. Lyuboshitz and
S. Pratt for the discussions on the sum rule presented here.}

\vfill
\eject
\vskip0.4in

\centerline{\bf References}

\vskip0.2in

\item{[1]} D. H. Boal, C.K. Gelbke and B.K. Jennings, Rev. Mod. Phys.
{\bf 62} (1990) 553.

\item{[2]} B. L\" orstad, Int. J. Mod. Phys. {\bf A4} (1989) 2861.

\item{[3]} St. Mr\' owczy\' nski, in Proceedings of International
Workshop on Multi-Particle Correlations and Nuclear Reactions
``Corinne II", Nantes, September 5 - 9, 1994, in print.

\item{[4]}  G.I. Kopylov and M.I. Podgoretsky, Yad. Fiz. {\bf 18} (1974) 656
(Sov. J. Nucl. Phys. {\bf 18} (1974) 336); {\it ibid.} {\bf 19} (1974) 434
({\bf 19} (1974) 215); G. Cocconi, Phys. Lett. {\bf B49} (1977) 459;
G.I. Kopylov, Phys. Lett. {\bf B50} (1974) 412;
S.E. Koonin, Phys. Lett. {\bf B70} (1977) 43;
M. Gyulassy, S. Kauffmann and L.W. Wilson, Phys. Rev. {\bf C20} (1979) 2267;
R. Lednicky and V.L. Lyuboshitz, Yad. Fiz. {\bf 35} (1982) 1316
(Sov. J. Nucl. Phys. {\bf 35} (1982) 770);
S.Pratt, Phys. Rev. Lett. {\bf 53} (1984) 1219.

\item{[5]} S. Pratt, Phys. Lett. {\bf B301} (1993) 159.

\item{[6]} Y.D. Kim, R.T. de Souza, C.K. Gelbke, W.G. Gong,
and S. Pratt, Phys. Rev. {\bf C45} (1992) 387;
B. Erazmus, L. Martin, and R. Lednicky, Phys. Rev. {\bf C49} (1994) 349;
R. Lednicky, V.L. Lyuboshitz, B. Erazmus, and D. Nouais, in Proceedings
of International Workshop on Multi-Particle Correlations and Nuclear
Reactions ``Corinne II", Nantes, September 5 - 9, 1994, in print.

\item{[7]} H. Sato and K. Yazaki, Phys. Lett. {\bf B98} (1981) 153;
E. Remler, Ann. Phys. {\bf 136} (1981) 293;
St. Mr\' owczy\' nski, J. Phys. {\bf G13} (1987) 1089;
V.L. Lyuboshitz, Yad. Fiz. {\bf 48} (1988) 1501
(Sov. J. Nucl. Phys. {\bf 48} (1988) 956);
St. Mr\' owczy\' nski, Phys. Lett. {\bf B248} (1990) 459;
{\it ibid.} {\bf B277} (1992) 43;
P. Danielewicz and P. Schuck, Phys. Lett. {\bf B274} (1992) 268.

\item{[8]} B. Jakobsson {\it et al}., Phys. Rev. {\bf C44} (1991) R1238.

\item{[9]} M.I. Podgoretsky, Yad. Fiz. {\bf 54} (1991) 1461.

\end